\documentclass[onecolumn,aps,preprint,showpacs,amsmath,amssymb]{revtex4}

\usepackage{graphicx}
\usepackage{dcolumn}
\usepackage{bm}

\begin{document}
\newcommand{\eq}{\begin{equation}}
\newcommand{\eqe}{\end{equation}}


\title{Angular distribution in two-photon double ionization of
helium \\ by intense attosecond soft X-ray pulses}

\author{Imre F. Barna$^{1,2}$, Jianyi Wang$^{1,3}$,  and Joachim Burgd\"orfer$^1$}

\affiliation{
$^1$ Institute for Theoretical Physics, Vienna University of Technology, A1040
Vienna Austria, EU, \\
$^2$Radiation and Environmental Physics Department, KFKI Atomic Energy Research Institute, P.O. Box
49, H-1525 Budapest, Hungary, EU,\\
$^3$ Department of Physics, University of Massachusetts Dartmouth,
North Dartmouth, MA 02747, USA}

\date{\today}

\begin{abstract}
We investigate two-photon double ionization of helium by
intense ($\approx 10^{15}$ W/cm$^2$) ultrashort ($\approx 300$ as)
soft X-ray pulses (E = 91.6 eV).
The time-dependent two-electron Schr\"odinger equation is
solved using a coupled channel method. We show that for ultrashort
 pulses the angular distribution of ejected electrons depends on the pulse
 duration and provides novel insights into the role of electron correlations
 in the two-electron photoemission process. The angular distribution at
 energies near the ``independent electron'' peaks is close to
 dipolar while it acquires in the ``valley'' of correlated emission
 a significant quadrupolar component within a few hundred attoseconds.

\end{abstract}

\maketitle

\section{Introduction}
Recent advances in the high-order harmonic generation (HHG)
techniques have led to
the development of soft X-ray sources that feature ultrashort pulses
 with pulse durations of a few hundred attoseconds (as) \cite{kie}
 and may reach intensities ($ \gtrsim 10^{14}$ W/cm$^2$) that
 are capable of inducing multiphoton processes. Extreme ultraviolet (XUV)
 pulses (photon energy 27.9 eV) with pulse duration of 950
 as have been characterized with an autocorrelation technique \cite{seki}.
 Recently, the two-photon double-ionization and above-threshold
 ionization of helium were experimentally observed with the Ti:sapphire
 27th harmonic pulses (photon energy 41.8 eV) \cite{hase}.
 These experimental advances open up the opportunity to revisit the dynamics
 of double ionization of helium by XUV photons previously investigated only
 in the single-photon absorption and scattering regime using synchrotron
 radiation \cite{sam,lev}. Simultaneous ejection of two electrons
 by a single photon allowed detailed tests of wavefunctions for the
 three-body Coulomb problem \cite{dor,mey,qiu} and the role of
 electron correlations in strongly inelastic processes accompanied
 with near-zero momentum transfer (photoabsorption) or
 sizable momentum transfer (Compton scattering) \cite{spiel,and}.

Multi-photon, in particular two-photon, ionization of helium
 by XUV pulses has been studied theoretically by different groups.
 A considerable numerical effort has been made to solve the two-active
 electron time-dependent Schr\"odinger equation (TDSE) with various methods.
 The R-matrix Floquet theory was successful to describe the
 (2$\gamma$,2e) process of He \cite{feng} in the photon energy range between
 where absorption of two photons are necessary for double ionization.
The configuration interaction B-spline spectral method \cite{lul,lul1}
was applied to solve the TDSE for this problem. The products of two
 B-splines represent the radial part of the wavefunction
which allows the inclusion of the electron-electron interaction
to a high degree of accuracy. Colgan {\it{et al.}} \cite{col}
 developed a time-dependent coupled channel method and
 studied the complete fragmentation
 of helium at 45 eV photon energy and presented fully differential
 cross sections.  Recently Lambropoulos {\it et al.} \cite{lam}
 found a ``knee'' structure in the intensity dependence reminiscent
 of a similar knee shape for double ionization by strong
 IR pulses \cite{walk}. Photons above the double ionization threshold
 $(\omega_{\mbox{xuv}} > 2.9 $ a.u.\ or 79  eV) were considered by
 Parker {\em et al.} \cite{park} who performed the direct numerical integration
 of the two-electron TDSE with a mixed finite-difference/basis
 set approach on a lattice and studied double-ionization with
 87  eV photon energy pulses with a laser peak intensity around
 10$^{16}$  W/cm$^2$. They analyzed both sequential as well as
 non-sequential double ionization events by a varying number of
 absorbed photons for long pulses ($\tau_p \gtrsim 2$   fs).
 Most recently, Ishikawa and Midorikawa \cite{ishi} investigated
 two-photon double ionization by ultrashort pulses with durations of
 $\tau_p \approx 150 $ to $450 $ as pertinent to HHG sources.
 They identified an ``anomalous'' component in the electron spectrum
 in between the two peaks associated with sequential double ionization and
 discussed its possible origin in terms of post-ionization energy
 exchange and core relaxation effects.

 In this paper, we theoretically
 investigate two-photon double ionization of helium by ultrashort attosecond
 pulses as a function of time by solving the TDSE with our coupled channel
 method which has been originally developed for heavy-ion helium
 collisions \cite{bar1,bar2,bar22} and later implemented to describe
 laser-driven atomic processes
 and two-photon
 coherent control \cite{bar3}.
 We consider experimentally realized
 high intensity laser pulses with 13.5 nm wavelength \cite{taka}
 which are the 59th harmonic of a Ti:sapphire laser (wavelength 800 nm).
 The photon energy considered (91.6 eV) is larger than the double ionization
 threshold of He (79 eV). A single photon is thus sufficient
 to induce double ionization. This case has been studied in detail
 with weak-field synchrotron sources where multi-photon effects are absent.
 Because one photon can interact with one electron only, double ionization
 cannot occur without electron-electron interaction. The picture is that
 one electron is directly ionized by absorbing the photon, and
 the second electron leaves through electron correlation either in the initial
 or in the final state, or both. This has been discussed in terms of
 a shake-off and electron-electron scattering
 (often referred to as TS1 \cite{mcgui97}).

 In contrast, for intense fields considered here, there is
 sufficient photon flux such that two photons can be absorbed,
 one by each electron, and the pair of electrons is ejected.
 Electron correlation is therefore not a prerequisite for double ionization
 to occur. At the same time, the ponderomotive energy of the
 XUV pulse $E_P = 8\pi I/4c \omega^2_{\mbox{xuv}}$ is so small that ionization
 by the rescattering of the first ionized and accelerated electron
 that causes ``non-sequential'' double ionization by strong IR pulses
 \cite{cork,web} can be ruled out.
 We discuss the conceptual difficulties in applying notions of
 sequential and non-sequential double ionization to such short pulses.
 We show that the angular distribution
 provides detailed insights into the ionization process on the attosecond time
 scale.  The role electron correlation plays in this process can be identified.

\section{Scenario for two-photon double ionization by attosecond XUV
 pulses: Time sequence and correlation}
We consider a linearly polarized attosecond XUV pulse with a Gaussian
 envelope,
\begin{equation}
\label{eq:1}
F (t) = F_0 \exp \left[ - 2 \ln 2\, \frac{t^2}{\tau_p^2}\right] \cos
(\omega_{\mbox{xuv}}t){\bf e}_z \, ,
\end{equation}
where $\tau_p$ is the full width at half-maximum (FWHM) of the pulse
 intensity.
The center frequency $\omega_{\mbox{xuv}}$ corresponds to a Ti:sapphire 59th
harmonic pulse with the energy of
91.6 eV. Following Ref. \cite{ishi} we will consider pulse durations
$\tau_p$ = 150 as and 450 as corresponding to $\tau_p $ = 6.25 and $\tau_p$
= 18.7 a.u. The period of the XUV cycle is $T = 2 \pi / \omega_{\mbox{xuv}} \cong $
1.9 a.u. The XUV pulse (Eq. \ref{eq:1}) subtends only few cycles
(3 to 10) and therefore closely resembles few-cycle optical or near-IR
pulses. The significant Fourier broadening therefore precludes the
appearance of spectrally sharp photoionization peaks. There is, however, a
fundamental difference to optical pulses of the same intensity: even at an
intensity of $ I = 10^{15}$ W/cm$^2$, the quiver amplitude of a free
electron, $  \sqrt{8\pi I/c} / \omega^2_{\mbox{xuv}}
 \approx 0.01 $ a.u.\ is small on
an atomic scale. Likewise, the ponderomotive energy
 $E_P = 8\pi I/4c \omega_{\mbox{xuv}}^2
\approx 0.0006$ a.u.\ 
is negligibly small. Therefore, ionization takes place deep in
the (multi) photon regime rather than in the tunnel ionization regime
applicable to IR pulses of the same intensity. This difference has immediate
consequences for the notion of ``(non) sequential'' ionization. While for
tunnel ionization the time window $\Delta t$ of an individual ionization
``burst'' can be uniquely identified near the field maxima with sub-cycle
precision \cite{lind}, the multi-photon ionization event is intrinsically
delocalized in time over several cycles. Only then does the electron
response to an electromagnetic pulse mimic that of photon absorption. In
view of the fact that the entire XUV pulse duration $\tau_p$ subtends only
a few cycles, it is obvious that the notion of sequentiality of ionization
events loses its meaning in the present case. This is in sharp contrast to
intense field ionization by optical fields. There, the first ionization by
tunnel  ionization under the influence of a quasi-classical electric field
is well localized and separated in time from the collisional ionization of
the second electron upon rescattering. The observed scaling with the pulse
duration $ \propto  \tau_p^N$, where $N$ is the number of photons absorbed,
should therefore not be taken as evidence of (non) sequentiality but a
measure of the total energy absorbed from the radiation field during
$\tau_p$.
The uncertainty in time when the absorption process takes place or time
delocalization of the multi-photon processes
does not imply that all time-differential information on the
ionization process is averaged out, as will be shown below.

The time characterizing the pulse duration should be compared to the time
scale of the electronic motion. Using the approximate hydrogenic expression
for the classical orbital period
\begin{equation}
\label{eq:2}
\tau_0 = 2 \pi n^3/Z^2_{eff} \, ,
\end{equation}
the orbital period ranges between $\tau_0 = 40 $ as (= 1.6  a.u.) for the
``inner'' electron of He$^+ (1s) (Z_{eff} = 2)$ and, for the ``outer'' electron
of He (1$s^2$) with a binding energy of 24.6 eV (Z$_{eff}$ = 1.3), $\tau_0$
= 90 as ($\approx$ 4 a.u.). The cycle period $T$ and the orbital period are
comparable to each other, thus probing the electronic motion on the
time scale on which the two interacting electrons of the helium ground state
exchange energy, linear and angular momentum. Thus, double ionization by
attosecond XUV pulses may probe electron correlations in both initial
and final states.

The role of correlation in double photoionization of helium is a
well-established subject in the low-intensity or single-photon limit of XUV
radiation going back to the pioneering paper by Byron and Joachain
\cite{byr}. As the electron-photon interaction is a one-body operator,
single-photon absorption can directly eject only one electron. Ejecting a
second electron requires with necessity electron-electron interaction. The
latter does not, however, imply correlation effects. Adhering here and in
the following to the identification of correlation with those pieces of the
interaction not included in a mean-field or independent particle (IP) model
as embodied in the (single configuration) Hartree-Fock description
\cite{mcgui97}, one-photon double ionization can proceed via mean-field
contribution. Already the sudden change of the screening following the
ejection of the first electron generates a finite probability of ejecting a
second electron. This ``shake-off'' process accounts for about 40 \% of the
total doble ionization cross sections at high photon energies. Clearly, for
a quantitatively accurate description, in particular over the entire range of
photon energies from threshold to high energies \cite{qiu} correlation
effects beyond the mean field in both the initial and
final states are essential.

\begin{figure}[ht]
\scalebox{0.5}{\rotatebox{-90}
{\includegraphics{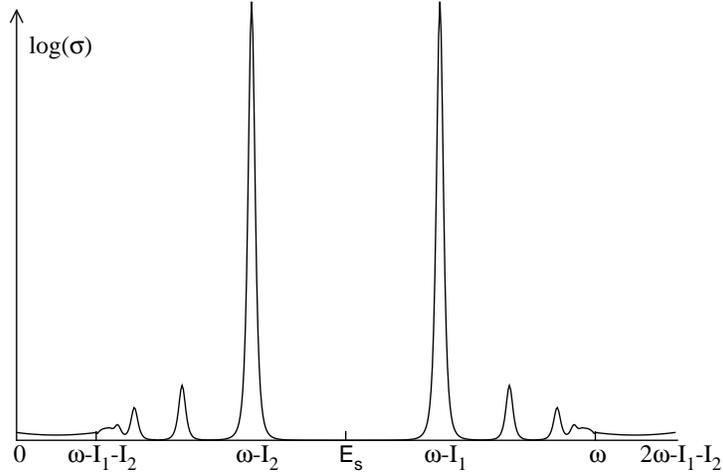}}}
\caption{\label{fig:fig1}  Electron spectrum following two-photon
absorption ($\omega\equiv\omega_{\mbox{xuv}}=$ 91.6 eV) in coincidence
with He$^{2+}$, schematically.
The symmetric energy is $E_S=\omega_{\mbox{xuv}}-(I_1+I_2)/2$ with
$I_{1,2}$ being the first and second ionization potentials, respectively.
}
\end{figure}

For two-photon double ionization by XUV pulses
 with $\omega_{\mbox{xuv}} > 2 $ a.u.\
dominance of independent particle (IP) ionization is expected since each
reaction
\eq
 \text{He} + \hbar \omega_{\mbox{xuv}} \rightarrow \text{He}^+(nl) + e^-,
 \label{single}
\eqe
\eq
 \text{He}^+(nl) + \hbar \omega_{\mbox{xuv}} \rightarrow \text{He}^{++} + e^-,
 \label{double}
\eqe
where $n$ and $l$ are the principal and angular momentum
 quantum numbers of He$^+$,
 respectively, is energetically allowed for all $n$.
 The quantum numbers, $n = 1$ and $l = 0$ are expected to dominate
 in Eqs. (\ref{single},\ref{double}).
Thus, correlation effects appear to be unimportant for the two-photon
process. It should be noted that the picture of a chain of reactions each
satisfying energy conservation in the photoelectric effect separately,
invoked by Eqs. (\ref{single}, \ref{double}) is only meaningful for
$\tau_p \rightarrow \infty$. In this limit, Eqs. (\ref{single},\ref{double})
implies an electron spectrum in coincidence with $\text{He}^{++}$ (displayed
schematically in Fig.\ 1) with two Rydberg series symmetrically centered
around the energy
\begin{equation}
\label{eq:5}
E_S = \omega_{\mbox{xuv}} - (I_1+I_2)/2 = 52 \, \mbox{eV}
\end{equation}
The single-photon double ionization spectrum well-known from
synchrotron-studies appears as a continuum below 12.6 eV
 ($\omega_{\mbox{xuv}} - I_1-I_2)$.  Its two-photon
replica would set in above $E=\omega_{\mbox{xuv}}=91.6$ eV.
 For ultrashort $\tau_p$ all discrete
peaks get dramatically broadened and merge into a quasi-continuum.

Apart from the broadening, the limit of short $\tau_p$ has further
consequences when this time becomes comparable to the electronic correlation
time $\tau_C$ in the helium ground state which can be simply estimated from
the characteristic time for collisional exchange of energy and angular
momentum between two classical electrons. Alternatively, it can be estimated
from the correlation energy $E_C = E - E_{HF}$ as $\tau_C = 1/E_C$. In
either cases, $\tau_{C}$ is of the order 10 a.u.\ (or 200 as).
XUV pulses with periods $T$ of 2 a.u.\ and durations of 3 - 10 a.u.\ therefore
 can probe the correlation dynamics.

\begin{figure}[ht]
\scalebox{0.5}{\rotatebox{-00}
{\includegraphics{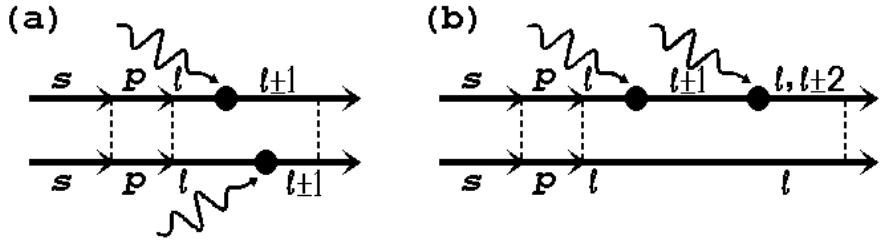}}}
\caption{\label{fig:fig2}  Schematic interaction diagrams for two-photon
absorption from He ground state.
(a) Each electron absorbs one photon each; (b) one
electron absorbs two photons.
Dashed lines denote the electron-electron interactions.
}
\end{figure}

It is instructive to visualize the two-photon double ionization process
diagrammatically (Fig.\ 2). The two photon lines each representing the
one-body operator of photoabsorption end either at the same or at two
different electrons resulting in two different diagrams. (The line
representing the nucleus has been omitted for simplicity). It should be
noted that a definite time ordering of the vertices of electron-photon
interactions is neither implied non meaningful for ultrashort pulses in
light of the discussion above. The dashed lines refer to electron-electron
interaction in the initial and final states which lead to energy and angular
momentum exchange. The latter is reflected in a configuration-interaction
wavefunction in terms of admixtures of orbitals of different
single-particle angular momenta,
\begin{eqnarray}
\label{eq:6}
|\Psi_i \rangle & = & \sum_{i,j} a_s^{(i,j)} |s^i \rangle | s^j \rangle +
\sum_{i,j} a_p^{(i,j)} |p^i \rangle | p^j \rangle\\
\nonumber
&& + \sum_{i,j} a_d^{(i,j)} |d^i \rangle | d^j \rangle + \dots
\end{eqnarray}
Typical orders of magnitude of admixture coefficients
for the initial state are \cite{byr}  (see also Eq.\ (\ref{he:func}) below)
 $|a_p/a_s|  \lesssim 0.1, |a_d/a_s| \lesssim 0.01$ and
those of higher $l$ are exceedingly small. The admixture of non-$s$ orbitals
to the He ground state provides a unique signature of electron correlation
as it would be absent in an IP or HF model.
More precisely, $l \ne 0$ configurations represent angular
correlation while coefficients $a_s^{(i,j)}$ may contain radial correlation.
Correspondingly, angular momentum
components in the final state reflect both the angular momentum transfer
$\Delta l = \pm 1$ by photoabsorption as well as the non-$s$ admixtures due to
electron correlations in the initial and final state. Their presence can be
mapped out by the time-dependence of the angular distribution of ejected
electrons.

\section{Computational method}
We have calculated the double ionization by two-photon absorption
represented by the diagrams of Fig.\ 2 using our time-dependent
coupled-channel  method. The point to be noted is that while we
discuss and interpret our results within to lowest-order perturbation theory
(LOPT), the calculation is fully non-perturbative taking into account
electron-photon and electron-electron interactions to all orders, albeit
within a truncated basis. Briefly, we calculate ionization process of the
helium atom in the laser pulse by solving the TDSE equation
\eq
 i\frac{\partial }{\partial t}\Psi({\bf{r}}_1,{\bf{r}}_2,t)=
 \left(\hat{H}_{He}+ \hat{V}(t)  \right)\Psi({\bf{r}}_1,{\bf{r}}_2,t),
 \label{Schtime}
 \eqe
 for the atomic Hamiltonian,
 \eq
 \hat{H}_{He} = \frac{{\bf{p}}_1^2}{2} +\frac{{\bf{p}}_2^2}{2} - \frac{2}{r_1}
 -\frac{2}{r_2} + \frac{1}{|\bf{r}_1 -\bf{r}_2|},
 \label{hamil}
 \eqe
and with the laser-electron interaction
 \eq
 \label{eq:9}
 \hat{V}(t) = - \sum\limits_{i=1,2}{\bf{F}}(t)\cdot {\bf{r}}_i
 \eqe
in the length gauge and the dipole approximation. The laser pulses are
linearly polarized along the $z$-axis with the time dependence given by Eq.
(\ref{eq:1}). We expand $\Psi({\bf{r}}_1,{\bf{r}}_2,t)$ in the basis
  $\{\Phi_{j}\}$ of eigenfunctions of the time-independent Schr\"odinger
  equation
  \eq
  \hat H \Phi_j({\bf{r}}_1,{\bf{r}}_2)=E_{j} \Phi_j({\bf{r}}_1,{\bf{r}}_2),
  \label{TISE}
  \eqe
  to yield
  \eq
  \Psi({\bf{r}}_1,{\bf{r}}_2,t) =\sum\limits_{j=1}^{N}a_j(t)
   \Phi_j({\bf{r}}_1,{\bf{r}}_2)e^{-iE_jt}\,,
    \label{Psi}
    \eqe
    where the $a_j(t)$ are the time-dependent expansion
    coefficients  and $E_{j}$ are the eigenvalues of (Eq. \ref{TISE}).
    Inserting (Eq. \ref{Psi}) into the TDSE
     (Eq. \ref{Schtime}) leads to the system of first-order
     differential equations for the expansion coefficients
     \eq
     \frac{da_k(t)}{dt} = -i\sum\limits_{j=1}^{N} V_{kj}e^{i(E_k-E_j)t}
     a_j(t)  \quad (k=1,...,N).
     \label{kopequ}
     \eqe
     Denoting the ground state by $k=1$, we impose the  initial condition
     \eq
     \label{eq:13}
      \begin{aligned}
      a_{k} \left( t \rightarrow - \infty \right)       \end{aligned}
     \left\{
     \begin{aligned}
     & 1  \quad &  k=1   \\
     & 0  \quad &  k\neq1.
     \end{aligned}
     \right.
     \eqe
     The asymptotic probabilities for transitions into final states $k$
after the      pulse has been turned off are given by
     \eq
     \label{eq:14}
     P_k = |a_k(t \rightarrow +\infty)|^2.
     \eqe
     The ionization probability can be retrieved from $P_{k}$ which
     includes discretized channels representing the continuum formed by the
wave packets.
      The equations of coupled channels (Eq. \ref{kopequ}) are solved by a
Runge-Kutta-Fehlberg integrator of order five with automatic time step
adjustment.

      The eigenfunctions $\Phi_{j}$ in (Eq. \ref{TISE}) are obtained by
      diagonalizing the Hamiltonian
      in a basis of orthogonal
       symmetrized two-particle functions
       $f_{\mu}$
       \eq
       \Phi_j({\bf{r}}_1,{\bf{r}}_2) = \sum\limits_{\mu}
       b_{\mu}^{[j]}f_{\mu}({\bf{r}}_1,{\bf{r}}_2)\,.
       \label{he:func}
       \eqe
       In the following we
       restrict ourselves to singlet helium states only.
 The two-particle functions are made up of symmetrized
 single particle orbitals, $g_{\epsilon
l}(r) Y_l^m$, where the radial functions $g_{\epsilon l}$ consist of
radial Slater functions and radial regular Coulomb wave packets.
We note that the coefficients $b_{\mu}^{[j]}$
are related to the admixture coefficients
discussed earlier following Eq.\ (\ref{eq:6}).
The wave packets form a discrete representation of the Coulomb continuum
and can serve as building blocks of our finite basis \cite{bar1,bar2}.

We include single-particle wavefunctions  with $ 0 \le l_1,l_2 \le 2$
angular momenta and
couple them to $ 0 \le L \le 2 $ total angular momentum two-electron states.
For the $L=0$ configurations we use $ss+pp+dd$ angular correlated
wavefunctions, for
 $L=1$ we use $sp+pd$ couplings and for $L=2$ the $sd+pp+dd$ configurations,
respectively. Since already the contribution of $d$ orbitals in the present
case is found to be small, higher $l_i$ can be safely neglected. The angular
correlated contributions play an essential role to understanding
the angular distribution of the ionized electrons.
In order to determine the final electronic state population, the expectation
value of the reduced one-electron density operator
$ \hat \rho = \sum_{i=1,2} \delta(\bf{r} -\bf{r}_i)$ is calculated after the
laser pulse,
\begin{equation}
\label{eq:16}
\rho(\vec{r})=\langle \Psi \left( t \rightarrow \infty \right) |\hat{\rho}|\Psi \left(t + \infty \right) \rangle.
\end{equation}
We employ the Feshbach projection method \cite{bar2} to separate the
singly-ionized states from the doubly-ionized states. Accordingly, the
one-electron polar angular distribution of ionized electrons in the double
ionization channel is given by
\begin{eqnarray}
\label{eq:17}
P_{DI}(\theta) &=& \frac{1}{2\pi} \int\limits_{0}^{2\pi} \int\limits_0^{\infty}
\langle \Psi_{DI} | \sum_{i=1,2} \delta({\bf{r}}-{\bf{r}}_i) |
 \Psi_{DI}  \rangle r^2 dr d\varphi   \nonumber \\
 &=& \frac{1}{\pi}  \int\limits_{0}^{2\pi} \int\limits_0^{\infty}  \int\limits_{{\bf{r}}_1}
|\Psi_{DI}(r,\theta,\varphi; {\bf{r}}_1)  |^2  d{\bf{r}}_1 r^2 dr d\varphi.
\end{eqnarray}
where $\Psi_{DI}$ represents the projection of $\Psi$ onto the subspace of
doubly ionized states.

\section{Results and Discussion}

\begin{figure}[ht]
\scalebox{0.5}{\rotatebox{-90}
{\includegraphics{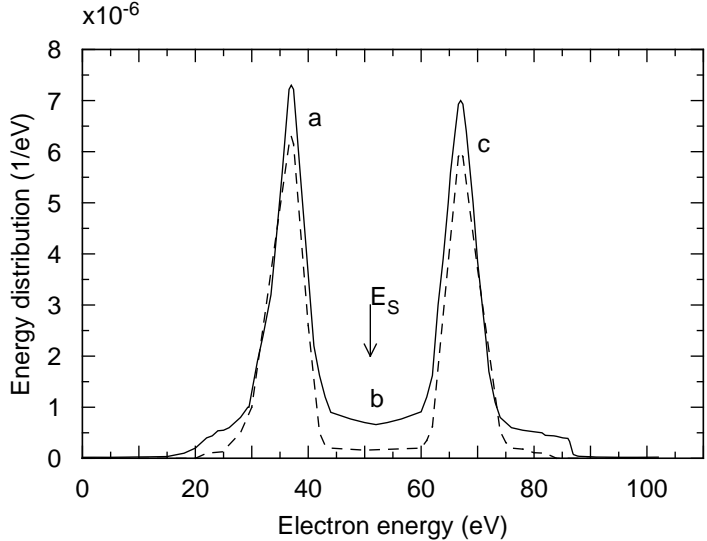}}}
\caption{\label{fig:fig3}
Energy distribution of the ejected
electrons in two-photon double-ionization of He.
The peak intensity of the pulse is 10$^{15}$ W/cm$^2$ and the
pulse duration is $\tau_p$=450 as.
The solid line represents our results and
the dashed line represents the data for Ishikawa {\it{et al.}} \cite{ishi}.
The energy positions referred to in text are 37.2 eV (a),
52 eV (b), and 67 eV (c).
}
\end{figure}

Before analyzing the angular distribution from Eq. (\ref{eq:17}) we briefly
present results for the energy distribution for which a direct comparison
with a recent calculation by Ishikawa and Midorikawa \cite{ishi} is
possible. The single electron energy distribution integrated over the second
electron for the pulse duration of 450 as (Fig.\ 3) features two prominent peaks
which can be easily identified with help of Fig.\ 1 as the ionization
spectra following the reactions Eqs. (\ref{single} and \ref{double}). The
dominant yet strongly broadened peak at 67 eV (labeled $c$) is due to
electrons ejected from the ground state of He with the first ionization
potential of $I_1$ = 24.6 eV. In the second interaction, the electrons are
ejected from
the He$^+$ ion with an ionization potential of $I_2 =$ 54.4 eV,
yielding the peak at $91.6-I_2=37.2$ eV (labeled $a$). From the higher members
or the Rydberg series only $n=2$ peaks are identifiable in Fig.\ 3 as local
humps, one
just below and one above the main peaks, respectively.
Structures from $n\ge 3$ are not visible since their
contributions become exceedingly small.
The cross section of the single-photon double ionization
continuum below 12.6 eV (see Fig.\ 1) is by far too small
to be visible on a linear scale.
The peaks $a$ and $c$ (Fig.\ 3) have been previously
referred to as sequential ionization \cite{ishi} or above-threshold
ionization \cite{park}.
We will refer to this process as independent
particle (IP) ionization to stress that electron correlation effects play no
significant role in their occurrence. This is in striking contrast to the
spectral feature in the ``valley'' (labeled $b$) (also referred to as
anomalous component \cite{ishi}) in which correlation effects are of crucial
importance. We refer to this feature as ``correlation induced'' (CI)
ionization.

\begin{figure}[ht]
\scalebox{0.5}{\rotatebox{-90}
{\includegraphics{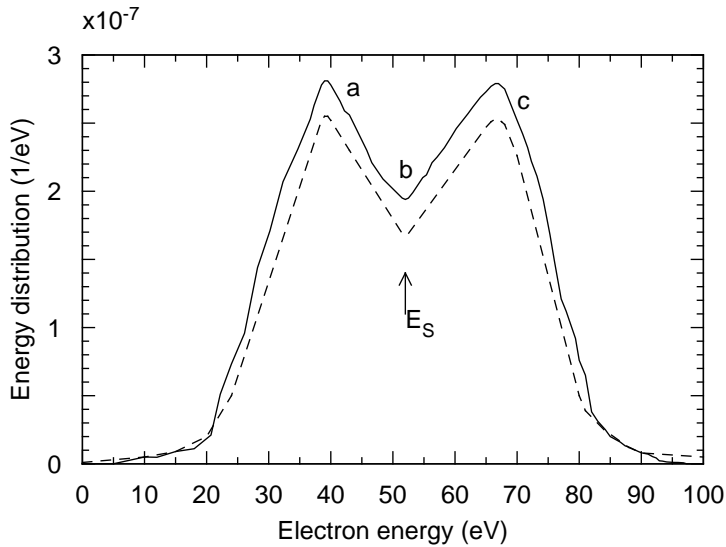}}}
\caption{\label{fig:fig4} Notation is the same as in Fig.\ 3,
 but for $\tau_p=150$ as.
The marked energy positions are 39 eV (a), 52 eV (b),
and 67 eV (c).
}
\end{figure}

The dependence of the CI ionization on the pulse duration $\tau_p$ is
illustrated in Fig.\ 4 for an ultrashort pulse of 150 as. The valley is now
quite shallow and a significant fraction of the ionization probability is
contained in the ``valley''. This is, in the first place, an obvious
consequence of the increased Fourier broadening in the ultrashort pulse limit.
In the opposite limit $\tau_p \gg T$ and $\tau_p \gg \tau_0$, the spectrum is
expected to revert to the quasi-discrete line spectrum, schematically
depicted in Fig.\ 1. Only in the long-pulse limit, the notions of time
ordering or sequentiality of the process takes on physical meaning. Overall,
our data agree with those of Ref. \cite{ishi} remarkably well on an absolute
scale with the largest discrepancies in the wings of the peak for the 10
cycle pulse (450 as).

\begin{figure}[ht]
\scalebox{0.4}{\rotatebox{-90}
{\includegraphics{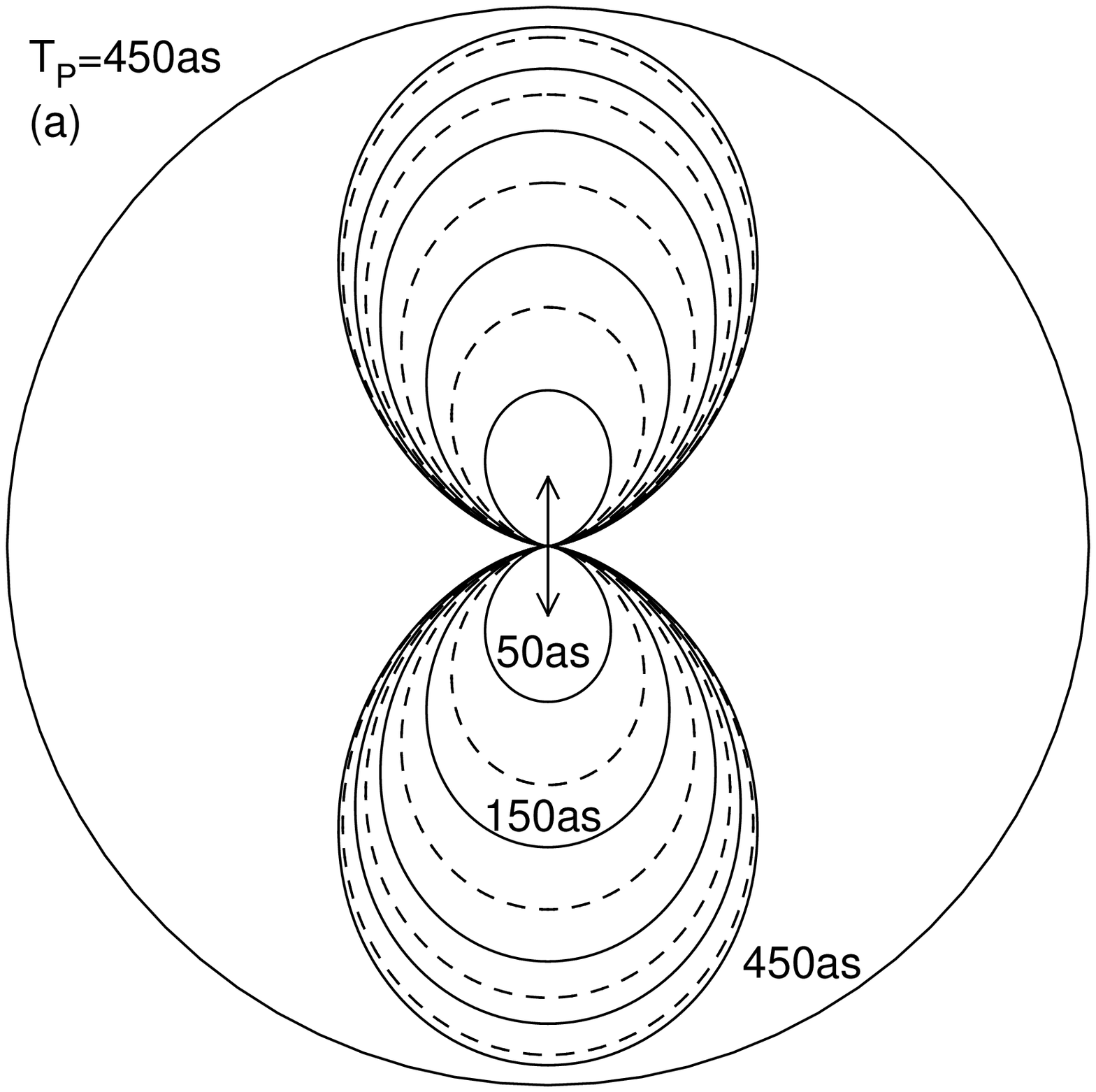}}}
\scalebox{0.4}{\rotatebox{-90}
{\includegraphics{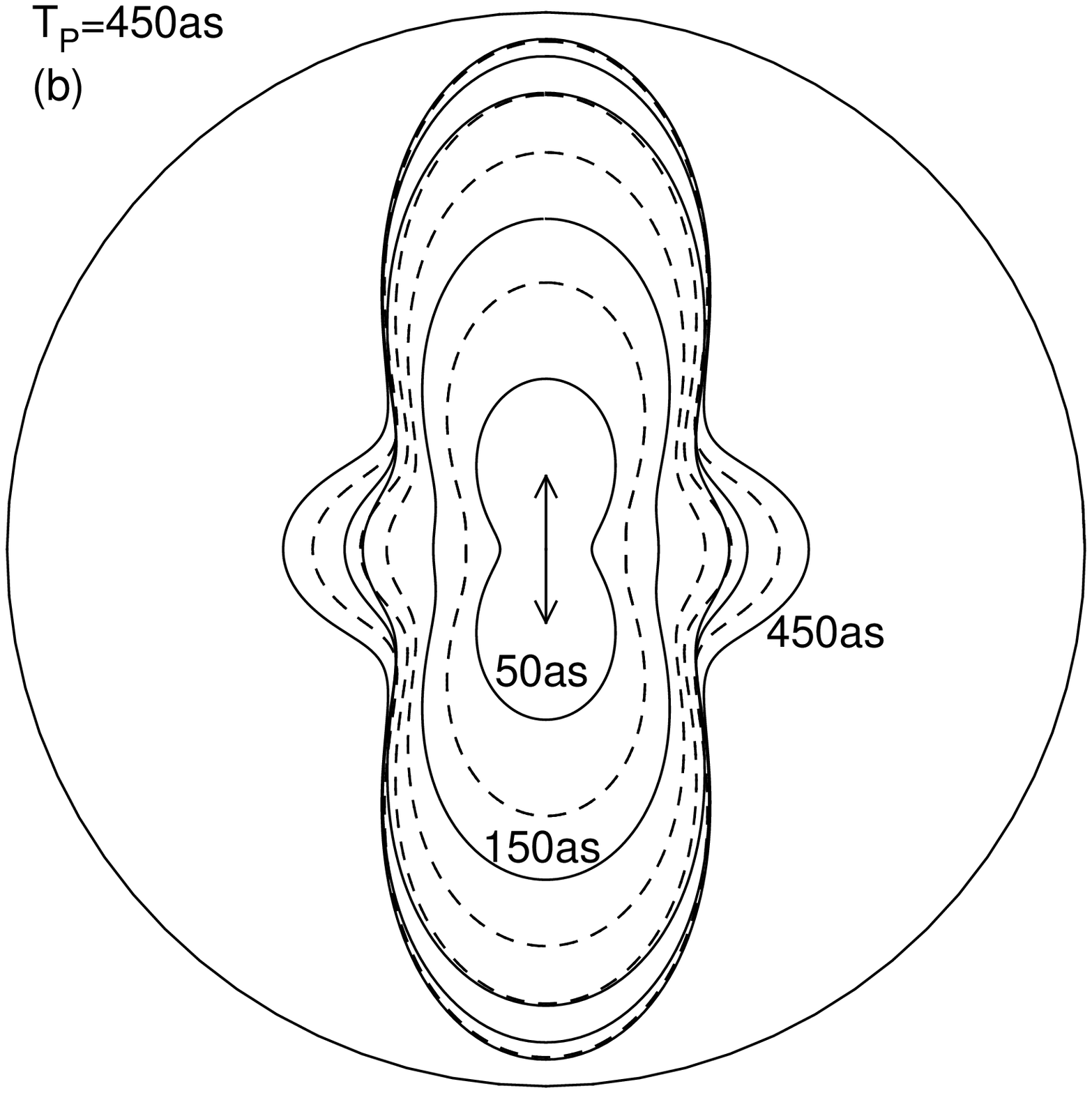}}}
\caption{\label{fig:fig5}
 The angular distribution (polar plot) of the ejected electrons for
an XUV pulse with $\tau_p=$ 450 as. Snap shots
of lines of constant intensities are taken
at times 50 to 450 in steps of 50 as (from inside going outward)
after the pulse's rise to half maximum,
for energies 37 eV (left) and 52 eV (right).
The unit circle indicates intensities of $7\times 10^{-6}$ eV$^{-1}$
for (a) and $6\times 10^{-7}$ eV$^{-1}$ for (b).
The arrows show the polarization axis.
}
\end{figure}

The identification of the valley region near $E_s$ with correlated
ionization is, in the first instance, taken over from one-photon double
ionization by synchrotron radiation where the region of symmetric energy
sharing of the available photon energy is dominated by correlation effects
\cite{mcgui97}. The extension of this identification to two-photon absorption
can be quantitatively justified by the properties of the angular distribution,
as shown below. Were the valley simply the result of the Fourier broadening
of two IP peaks, the distribution $P_{DI} (\theta)$ at the energies near
$(E_S)$ should closely resemble those of the spectral regions ($a$)  or ($c$).
That this is not at all the case is illustrated by
the polar plot (Fig.\ 5) of the
angular distribution near the IP ionization peak ($a$) and the CI ionization
valley ($b$). The different contour line indicates the time evolution of the
angular distribution in increments of 50 as for the 450 as pulse. While the
IP peak retains the emission pattern of
a Hertz dipole during the entire pulse duration, the CI
electron distribution takes on a pronounced non-dipolar, i.e. quadrupolar,
pattern after about 150 as. The onset of a non-dipolar distribution on this
timescale can be also observed for the ultrashort pulse of $\tau_p = 150$ as
(Fig.\ 6) indicating that the sharp differences in the angular distribution
between the IP peak and the CI component is also present when the valley is
very shallow. For the ultrashort pulse a slight peak shift from 37 eV to
about 39 eV is found in agreement with Ref. \cite{ishi}.

\begin{figure}[ht]
\scalebox{0.4}{\rotatebox{-90}
{\includegraphics{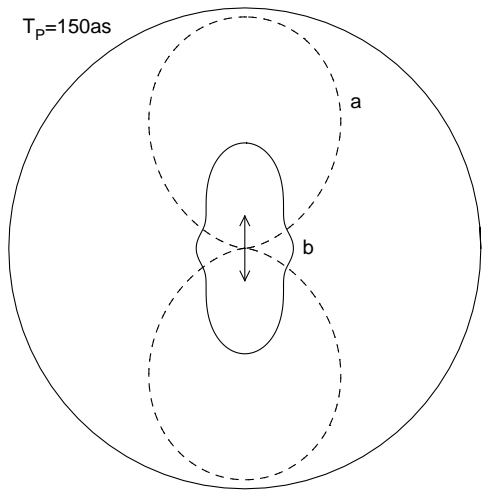}}}
\caption{\label{fig:fig6}
Polar plots of  the angular distribution  of the ejected electrons
after a pulse with $\tau_p=$ 150 as.
The distribution (a) is taken at 39 eV and (b) at 52 eV.
The unit circle indicates an intensity of $3\times 10^{-7}$ eV$^{-1}$.
The arrows show the polarization axis.
}
\end{figure}

The anisotropy of the angular distribution can be characterized by the
multipole expansion
\begin{equation}
  \frac{d\sigma}{d\Omega}=\frac{\sigma_0}{4\pi}
 \left [1+\beta P_2(\cos \theta) +\gamma P_4(\cos \theta) \right ],
\label{bega}
\end{equation}
where $\sigma_0$ is the integral cross section,
$P_{2,4}$ are the Legendre polynomials, and $\beta$ and
$\gamma$ are the second order ($k=2$) and fourth-order ($k=4$)
anisotropy parameters, respectively.
Note that a ``dipolar'' emission pattern has $k=2$, i.e., it
represents ``alignment'', while the ``quadrupolar'' pattern
is of rank $k=4$ and should be more correctly referred to
as ``hexadecapole''.
Individually, the range of the multipole parameters are
$-1\le \beta \le 2$, and $-1\le \gamma \le 7/3$, the highest order of
anisotropy $k = 4$ is consistent with two-photon absorption,
\begin{equation}
\label{eq:18}
k \leq 2N \, .
\end{equation}
Higher anisotropy coefficients beyond $k = 4$ are not detectable.
By projecting the numerically calculated angular distributions
to Eq.\ (\ref{bega}), we obtain
the $\beta$ and the $\gamma$ parameters listed in Table 1. Near the IP
ionization peaks, (37 and 39 eV for 450 and 150 as pulses, respectively)
$\beta$ is at least one order of magnitude larger than $\gamma$. The $\beta$
values are very close to their maximum value of 2.

At 52 eV, near the CI valley, $\beta$ and $\gamma$ become comparable, giving
rise to a strong mixing of dipole and quadrupole terms in the angular
distributions.

\begin{table}[ht]
\caption{\label{t2}
The multipole expansion parameters $\beta$ and $\gamma$ Eq.
(\ref{bega}) for the pulse durations of 450 and 150 as,
at two energies each, corresponding to IP and CI ionization (see text).}
\vspace{2ex}

\begin{tabular}{lrrrr} \hline \hline
$\tau_p$  & Type & Energy & $\beta$ & $\gamma$ \\
\hline
450 as & IP & 37 eV &1.94 &-0.08\\
       & CI & 52 eV   &0.40 & 0.58 \\
\cline{2-5}
150 as & IP & 39 eV & 1.87  & -0.17\\
       & CI      & 52 eV & 0.51  & 0.35  \\
\hline \hline
\end{tabular}
\end{table}

It should be noted that the present deviation from a strictly aligned
 ($k = 2$) pattern is due to multiphoton effects and not due to retardation
effects beyond the dipole approximation \cite{derev}.
Figs.\ 5 and 6 clearly show that two-photon IP ionization features a near
Hertz dipole distribution while CI ionization possess a significant $k = 4$
admixture. It is now instructive to relate the origin of the quadrupole
component to correlations. A non-vanishing $\gamma$ requires a final state in the
continuum with $L_f = 2$ since $k = 2 L_f$. The latter results from coupling
of configurations involving single-particle orbitals, $(l_f, l_{f}'): (s_f
d_f), (p_f p_f)$ and $(d_f d_f)$, where the latter is already negligible at
the present intensity. By selectively switching off final states consisting
of $(s_f d_f)$ and $(p_f p_f)$ configurations we find that the IP ionization
peak is dominated by $(p_f p_f)$ orbitals while the CI ionization
contribution is dominated by $(s_f d_f)$ contributions. These final states
can be reached by absorption of two photons along the
LOPT pathways that
correspond to either diagram (Fig.\ 2b)
\begin{subequations}
\label{eq:20}
\begin{align}
\label{eq:20a}
(s_i \rightarrow s_f, s_i \stackrel{\mbox{xuv}}{\to}
                          p   \stackrel{\mbox{xuv}}{\to}  d_f)
\end{align}
or diagram (Fig.\ 2a)
\begin{align}
\label{eq:20b}
(p_i \stackrel{\mbox{xuv}}{\to} s_f, p_i \stackrel{\mbox{xuv}}{\to} d_f)
\end{align}
While the first path (Eq. (\ref{eq:20a})) can be realized for the dominant
configuration in the initial state $(s_i, s_i)$ (see Eq. \ref{eq:5})) and
would be present for an uncorrelated initial state described by e.g. HF
wavefunction, the second path (\ref{eq:20b}) has as prerequisite
configuration admixtures $(p_i, p_i)$ to the initial state and thus
initial-state angular correlation. When selectively eliminating the $(p_i,
p_i)$ configuration from the  initial state we find that the cross section
in the valley region is reduced by almost an order of magnitude. This
unambiguously characterizes the ``anomalous'' cross section component in the
``valley'' as being due to correlations. By contrast, the IP ionization
peaks are barely affected when $(p_i p_i)$ configurations are removed. This
is plausible as the dominant two-photon absorption process from an
uncorrelated initial state according to Fig.\ 2a
\begin{align}
\label{eq:20c}
(s_i \stackrel{\mbox{xuv}}{\to} p_f, s_i \stackrel{\mbox{xuv}}{\to} p_f)
\end{align}
\end{subequations}
predicts a dominance of a Hertz dipole pattern for each ejected electron.
Our calculations suggests that initial-state correlations may be more
important than final-state correlations.
This is due to the fact that the pair of electrons near the symmetric energy
sharing point $E_s = 52$ eV leave the interaction region quickly with a
relatively large speed of $v=2$ a.u.

Ishikawa {\em et al.} \cite{ishi} have discussed the ``anomalous'' component in
terms of two semiclassical models. Post-ionization energy exchange (PIEE)
and second ionization during core relaxation (SICR). They found that PIEE is
inefficient to account for the valley region consistent with our observation
that final-state correlations are of minor importance. On the other hand the
relaxation process due to change in screening in the SICR appears to
resemble somewhat a shake process and is as a quasi-isotropic process
unlikely to yield a high-order ($k = 4$) anisotropy.

\section{Conclusions}
We have studied the electron energy and angular distributions
in two-photon double-ionization of He by an attosecond, intense soft X-ray
pulse, specifically, for the Ti:sapphire 59th harmonic pulse with an
 intensity of 10$^{15}$ W/cm$^2$.
We solved the TDSE with our coupled channel method in which
the electron-electron interaction is fully taken into account.

The electron energy distributions show well-localized peaks
for pulse of long duration $\tau_p$. They are understood to arise from
the independent particle (IP) ionization. For short pulses of only a few
hundred attoseconds, the peaks
shift toward each other and the cross section in the valley between the
peaks becomes significant.
We attribute this ionization component to the correlation-induced (CI)
ionization. We investigated the electron angular distributions from
IP and CI ionization.
We find shape profiles to be that of a Hertz dipole for
IP ionization but a significant admixture of a $k = 4$ (``quadrupole'')
components for CI ionization.
The unique signature of correlation-induced ionization
is the presence of this $k = 4$ component in the angular shape profiles.
They were further quantified in terms of the multipole expansion parameters.

Time evolution of the electron angular distribution suggests
that sequentiality of electron ejection or
photon absorption is neither relevant nor well-defined. Clearly, further
studies are needed to clarify
electron correlation effects.
Joint energy-angular distributions
(i.e., kinematically complete momentum distributions)
 would provide new insight into the ionization mechanism.
It would also be useful to understand the ionization
with the help of a perturbative approach, either with the
electron-electron interaction or the pulse intensity
as the expansion parameter. This would provide a
complementary picture to various mechanisms
that may be difficult to identify in fully numerical TDSE results.

\begin{acknowledgments}
We would like to thank Dr.\ Ishikawa for providing us
their unpublished results.
This work was supported by SFB 016-FWF and EU-HITRAP, Project
Number HPRI-CT-2001-50067.
\end{acknowledgments}


\end{document}